\documentclass{aa}

\usepackage[intlimits,fleqn]{amsmath}
\usepackage[mathscr]{eucal}
\usepackage[english]{babel}
\usepackage{graphicx}
\usepackage{txfonts}

\usepackage{natbib}
\bibpunct{(}{)}{;}{a}{}{,}

\newcommand*{\ud}{\mathrm{d}}

\begin{document}

\title{Fourier phase analysis in radio-interferometry}

\author{F. Levrier\inst{1}
\and E. Falgarone\inst{1}
\and F. Viallefond\inst{2}
}

\offprints{F. Levrier, \email{levrier@lra.ens.fr}}

\institute{LERMA - UMR 8112 du CNRS, LRA, \'Ecole normale sup\'erieure, 24 rue Lhomond, 75231 Paris Cedex 05, France
\and LERMA - UMR 8112 du CNRS, Observatoire de Paris, 61 Avenue de l'Observatoire, 75014 Paris, France
}

\date{}

\abstract{
Most statistical tools used to characterize the complex structures of the interstellar medium can be related to the power spectrum, and therefore to the Fourier amplitudes of the observed fields. To tap into the vast amount of information contained in the Fourier phases, one may consider the probability distribution function (PDF) of phase increments, and the related concepts of phase entropy and phase structure quantity. We use these ideas here with the purpose of assessing the ability of radio-interferometers to detect and recover this information. By comparing current arrays such as the VLA and Plateau de Bure to the future ALMA instrument, we show that the latter is definitely needed to achieve significant detection of phase structure, and that it will do so even in the presence of a fair amount of atmospheric phase fluctuations. We also show that ALMA will be able to recover the actual ``amount'' of phase structure in the noise-free case, if multiple configurations are used.

}

\maketitle

\section{Introduction}

The physics of the interstellar medium (ISM) stands at the crossroads of many astrophysical problems, from stellar formation to galaxy evolution.  Without a proper understanding of the processes taking place in the ISM, and of their interplay, complete and satisfactory solutions to these problems cannot hope to be met.

Turbulence is one such process \citep[see e.g.][]{spicker88,odell87,miesch94}, and it is thought to play a major role in the shaping of the fractal structures observed \citep{falgarone91,vogelaar94,elmegreen96,falgarone98a,stutzki98,elmegreen2001}.

Consequently, a quantitative description of these structures is a necessary first step towards understanding the physics of the ISM, and many statistical tools have been used to this end. Let us mention the power spectrum \citep[see e.g.][]{gautier92,dickey2001,stanimirovic2001}, the autocorrelation function \citep{kleiner85,perault86}, the $\Delta$-variance \citep{stutzki98,bensch2001}, the fractal dimension \citep{falgarone91} and the wavelet decomposition \citep{gill90}.

These various tools are not altogether independent from one another. By definition, the autocorrelation function is the Fourier transform of the power spectrum, to which the $\Delta$-variance and fractal dimension can also be related, albeit less directly \citep{stutzki98}. Finally, the $\Delta$-variance can be written as the variance of wavelet transform coefficients \citep{zielinsky99}. On the whole, it is then fair to say that all of these tools are connected, in one way or another, to the power spectrum, although some of them are of easier and more reliable use depending on the type of observation \citep{stutzki98,bensch2001}. 

Since the power spectrum is given by the squared amplitudes of Fourier components, it basically ignores any structural information that may be contained in the Fourier phases. Now, each Fourier component corresponds to a plane wave in direct space, with a given wave vector, amplitude and phase. The Fourier transform being linear, the observed structures are the result of the interaction between the various plane waves. Ignoring the phases when characterizing the structures is thus comparable to ignoring the interference phenomenon, and therefore marks a major loss in structural information. This has been confirmed by simple numerical experiments \citep{juvells91,coles2005}. 

In the experiment performed by \citet{coles2005}, the Fourier phases of a numerical simulation of galaxy clustering, which is a highly-structured field, are randomly reshuffled in Fourier space. The resulting field has lost most of the filamentary structure observed in the original image. This shows that it is in the Fourier-spatial distribution of the phases, and not in their values themselves, that most of the structural information must lie.

The importance of this information may be best estimated in the context of interferometry. Indeed, interferometers essentially measure some Fourier components of the observed structures, and thus theoretically provide direct access to their phases. With the forthcoming ALMA instrument, the capacity of interferometers to detect structure in the Fourier phases \emph{in real time} may be assessed. This is the purpose of this paper, which is organized as follows: Section \ref{sec_pfa} offers a summary of the Fourier phase analysis technique, whose numerical implementation is presented in section \ref{sec_psqip}. The main part of the paper, dealing with the application of these techniques to interferometric observations, is the topic of section \ref{sec_atio}. Finally, section \ref{sec_conc} gives a summary and conclusions.

\section{Fourier phase analysis}
\label{sec_pfa}

The importance of Fourier phases in terms of structure has been exploited in various studies concerning variations of the magnetic field in cometary plasmas and Solar wind \citep{polygiannakis95}, the large-scale clustering properties of the Universe \citep{scherrer91,chiang2000,watts2003,coles2005}, and Cosmic Microwave Background (CMB) maps \citep{coles2004}.

As we noted earlier, it is in the Fourier spatial distribution of phases that information should be sought. To quantify this, \citet{scherrer91} suggested considering the statistics of phase increments $\Delta_{\boldsymbol{\delta}}\phi(\boldsymbol{k})=\phi(\boldsymbol{k}+\boldsymbol{\delta})-\phi(\boldsymbol{k})$ between points separated by a given lag vector $\boldsymbol{\delta}$ in Fourier space. 

\begin{figure}[htbp]
\resizebox{0.85\hsize}{!}{
\includegraphics{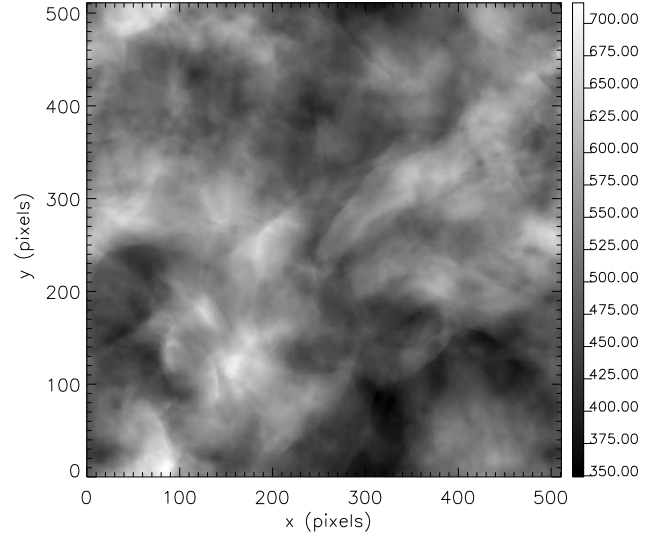}
}
\caption{Column density of a 512$^3$ weakly compressible hydrodynamical turbulence simulation obtained by \citet{porter94}, used here as a model brightness distribution for phase structure analysis.}
\label{fig_4}
\end{figure}

\begin{figure}[htbp]
\resizebox{\hsize}{!}{
\includegraphics{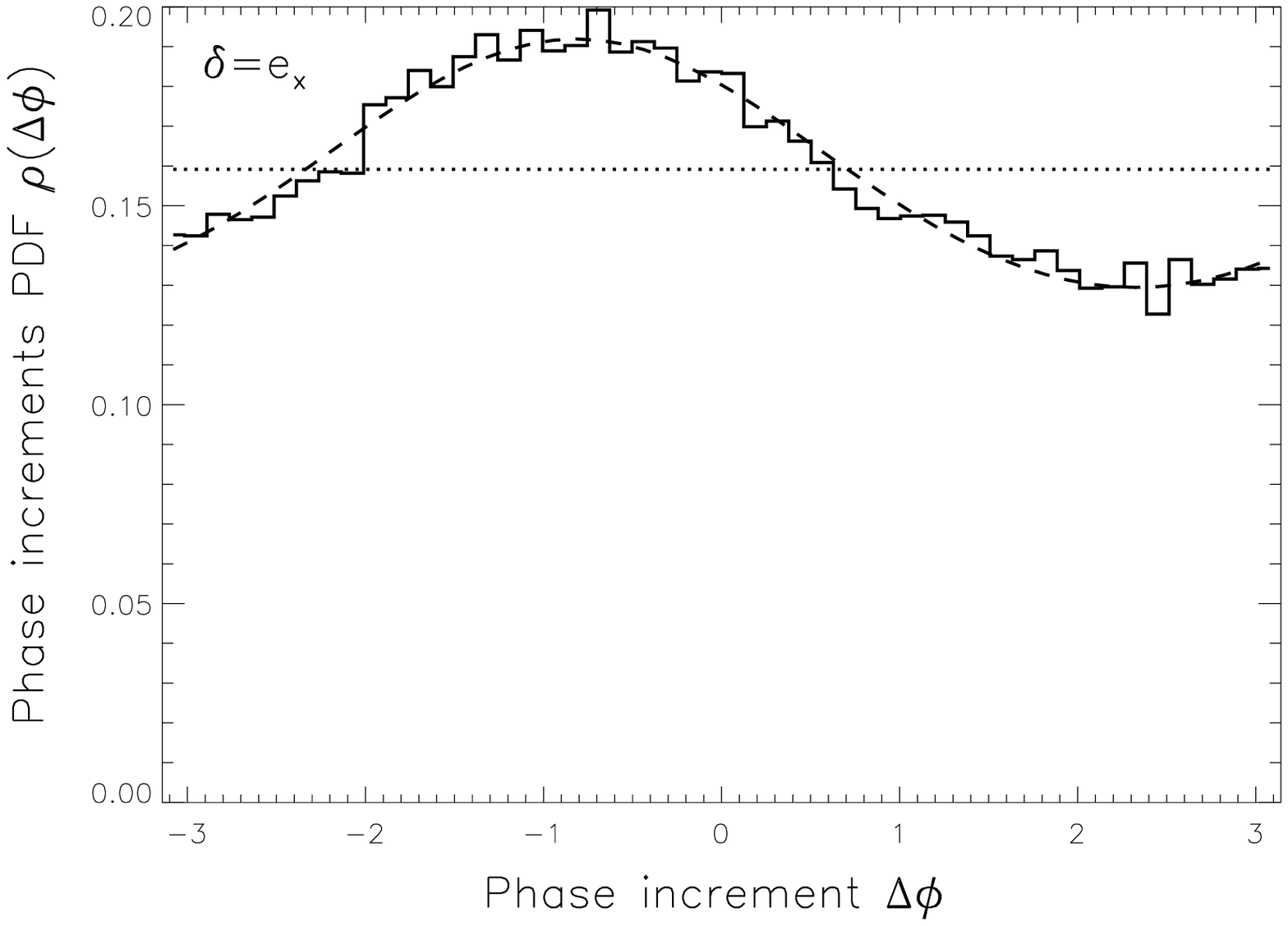}
}
\resizebox{\hsize}{!}{
\includegraphics{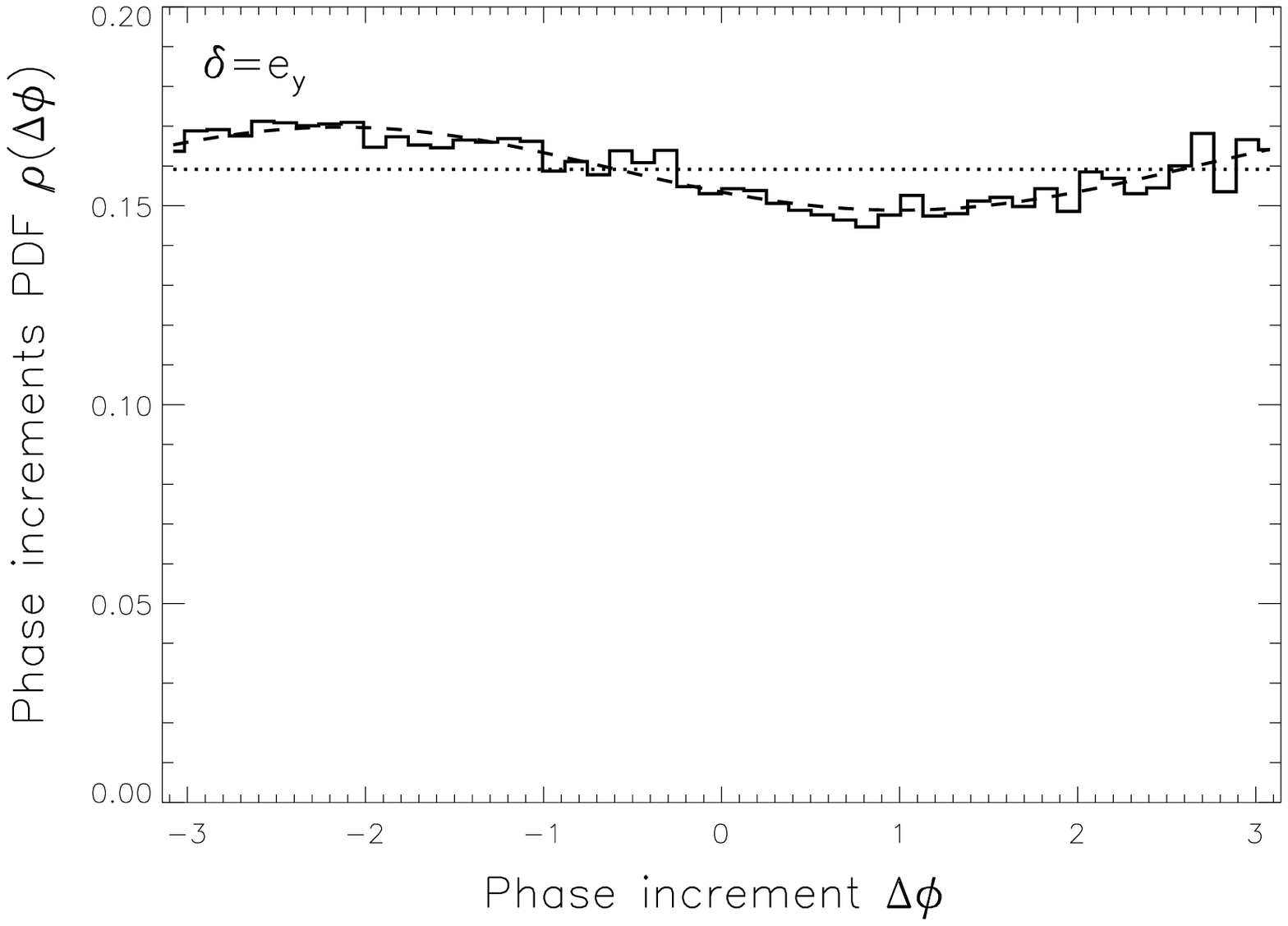}
}
\caption{Histograms of phase increments for the turbulent brightness distribution (Fig. \ref{fig_4}). The lag vectors used are $\boldsymbol{\delta}=\boldsymbol{e}_x$ (top panel) and $\boldsymbol{\delta}=\boldsymbol{e}_y$ (bottom panel). The dotted lines represent the uniform distribution and the dashed lines are fits by von Mises distributions (see text). The number of bins is $n=50$.}
\label{fig_5}
\end{figure}

In a field with no phase structure, all phases are uncorrelated, and phase increments are therefore uniformly distributed random variables. Conversely, if the probability distribution function (PDF) of phase increments deviates from uniformity, this deviation may be seen as a signature of phase structure. As an example, consider the column density of a $512^3$ weakly compressible hydrodynamical turbulence simulation \citep{porter94}, shown on Fig.~\ref{fig_4}. Since this field is periodic, its Fourier phases can be computed using a Fast Fourier Transform (FFT) algorithm. Given a lag vector $\boldsymbol{\delta}$, the PDF $\rho\left(\Delta_{\boldsymbol{\delta}}\phi\right)$ of phase increments for this lag is then approximated numerically by computing the histogram of $\Delta_{\boldsymbol{\delta}}\phi$ values. Two of these histograms are shown on Fig.~\ref{fig_5}, for lag vectors $\boldsymbol{e}_x$ and $\boldsymbol{e}_y$, which are the unit vectors\footnote{Their lengths are actually $1/N_x\Delta_x$ and $1/N_y\Delta_y$, where $\Delta_x\times\Delta_y$ is the actual size of a pixel in direct space and $N_x\times N_y$ is the image size in pixels.} of the Fourier space basis associated with a direct space basis $(\boldsymbol{u}_x,\boldsymbol{u}_y)$. We observe that these distributions are not uniform, with a single wavelike oscillation around the value $1/(2\pi)$, and that the amplitude of the oscillation is more important for $\boldsymbol{\delta}=\boldsymbol{e}_x$ than for $\boldsymbol{\delta}=\boldsymbol{e}_y$, which may be interpreted as evidence for anisotropy. 

For other lag vectors or different images, the shape of the histogram remains, although the amplitude may vary. As pointed out by \citet{watts2003}, the underlying distributions are very likely to be von Mises distributions, given by
\begin{equation*}
\rho\left(\Delta_{\boldsymbol{\delta}}\phi\right)=\frac{1}{2\pi I_0(\kappa_{\boldsymbol{\delta}})}\exp{\left[-\kappa_{\boldsymbol{\delta}}\cos{(\Delta_{\boldsymbol{\delta}}\phi-\mu_{\boldsymbol{\delta}})}\right]},
\end{equation*}
where $I_0$ is the zeroth order modified Bessel function of the first kind, and the parameters $\mu_{\boldsymbol{\delta}}$ and $\kappa_{\boldsymbol{\delta}} > 0$ control respectively the position of the distribution's minimum and the oscillation's amplitude. Thus, $\kappa_{\boldsymbol{\delta}}$ can be viewed as a measure of the amount of phase structure in the image. For the histograms of Fig. \ref{fig_5}, on which fits by von Mises distributions are shown, the values of $\kappa_{\boldsymbol{\delta}}$ found are $\kappa_{\boldsymbol{e}_x}=0.197$ and $\kappa_{\boldsymbol{e}_y}=0.0656$.

Phase entropy, introduced by \citet{polygiannakis95}, is another measure of the distribution's departure from uniformity, and is defined by the integral
\begin{equation*}
\mathcal{S}(\boldsymbol{\delta})=-\int\nolimits_{-\pi}^{\pi} \rho\left(\Delta_{\boldsymbol{\delta}}\phi\right)\ln{\left[\rho\left(\Delta_{\boldsymbol{\delta}}\phi\right)\right]}\ud\Delta_{\boldsymbol{\delta}}\phi.
\end{equation*}
It reaches its maximum value $\mathcal{S}_0=\ln{(2\pi)}$ for uniform PDFs, and tends to $-\infty$ for $\delta$-function PDFs. These extreme cases correspond respectively to fields with purely random phases such as fractional Brownian motions\footnote{These are random fields characterized by a power-law power spectrum and random phases.} \citep[see e.g.][]{stutzki98}, and to fields containing a single point-source. Given these limits, it is convenient to consider the positive quantity $\mathcal{Q}(\boldsymbol{\delta})=\mathcal{S}_0-\mathcal{S}(\boldsymbol{\delta})$, which we dub ``phase structure quantity'' in the rest of this paper. There is a monotonous relationship between the two measures $\kappa_{\boldsymbol{\delta}}$ and $\mathcal{Q}(\boldsymbol{\delta})$, given by\footnote{$I_1$ is the first order modified Bessel function of the first kind.}
\begin{equation*}
\mathcal{Q}(\boldsymbol{\delta})=\kappa_{\boldsymbol{\delta}}\frac{I_1(\kappa_{\boldsymbol{\delta}})}{I_0(\kappa_{\boldsymbol{\delta}})}-\ln{\left[I_0(\kappa_{\boldsymbol{\delta}})\right]},
\end{equation*}
but we shall deal only with the phase structure quantity $\mathcal{Q}$ from now on. The reason for this is computational: phase structure quantities can be computed on histogram-like functions, while the parameter $\kappa_{\boldsymbol{\delta}}$ requires fitting them by von Mises distributions first, a procedure which may not converge properly when the underlying distribution is close to uniform. For the von Mises distributions fitted on the histograms of Fig. \ref{fig_5}, $\mathcal{Q}(\boldsymbol{e}_x)=9.6\,10^{-3}$ and $\mathcal{Q}(\boldsymbol{e}_y)=1.1\,10^{-3}$, respectively. Using column density fields of the same compressible turbulence simulation at different times, we find that the typical values for phase structure quantities lie typically below $10^{-2}$, whereas \citet{chiang2000} found values as high as $\sim 0.4$ for gravitational evolution of density perturbations. Although density contrasts are quite different, making direct comparison somewhat hazardous, this suggests that Fourier phase analysis may be useful in determining the physical processes governing the formation of structures in the interstellar medium.

\section{Phase structure quantity in practice}
\label{sec_psqip}

Due to the limited number of phase increments that can be computed from a finite-sized image, the histograms do not perfectly sample the underlying PDFs, but rather include a certain amount of statistical noise. Consequently, the histograms of phase increments for images such as fractional Brownian motions are not exactly uniform, and the corresponding phase structure quantities are not zero. More generally, the phase structure quantities $\mathcal{Q}$ associated with the underlying distributions are to be distinguished from those found by numerical integration performed on the histograms, since these depend on the number $p$ of increments available and the number $n$ of bins used. We write $\tilde{\mathcal{Q}}$ for these estimates of phase structure quantities. For the histograms shown on Fig. \ref{fig_5}, built with $n=50$ and $p=512 \times 511$, we find $\tilde{\mathcal{Q}}(\boldsymbol{e}_x)=9.8\,10^{-3}$ and $\tilde{\mathcal{Q}}(\boldsymbol{e}_y)=1.3\,10^{-3}$, which shows that the difference with $\mathcal{Q}$ (about $2.10^{-4}$) can become significant for low phase structure quantities.

Assessing the detectability of phase structure makes it necessary to determine the threshold of $\tilde{\mathcal{Q}}$ above which it can be said that an image deviates significantly from a ``structureless'' field, given values for $n$ and $p$. Mathematically speaking, this amounts to determining an upper limit to the probability that the estimate $\tilde{\mathcal{Q}}$ of the phase structure quantity be greater than a given positive real number $x$, assuming a uniform parent distribution, as a function of $n$ and $p$.

We may obtain such an upper limit by an analytic approach, as described in detail in the appendix. The demonstration is based on results obtained by \citet{castellan2000} in her PhD thesis, which is available online\footnote{\tt \tiny http://www.math.u-psud.fr/theses-orsay/2000/6039.html}. These results are themselves derived from those of \citet{barron91} and show that an upper limit to the probability $\mathsf{P}\left(\left\{\tilde{\mathcal{Q}} > x\right\}\right)$ is given by the quantity
\begin{equation*}
n\left[1-\mathrm{Erf}\left(\epsilon\sqrt{\frac{p}{2(n-1)}}\right)\right]+\mathsf{P}\left(\left\{\chi^2 > \frac{2px(1-\epsilon)^2}{1+\epsilon}\right\}\right)=\mathsf{P}_1+\mathsf{P}_2,
\end{equation*}
where $\mathrm{Erf}$ is the error function and $\chi^2$ is the chi-square statistics of degree $n-1$. The first term $\mathsf{P}_1$ corresponds to the probability of having an histogram with a large fluctuation, the meaning of ``large'' being defined by means of the arbitrary positive real number $\epsilon$, as explained in the appendix. The second term $\mathsf{P}_2$ corresponds to the probability of having $\tilde{\mathcal{Q}} > x$ with a more regular histogram. Since the threshold value $x$ appears in $\mathsf{P}_2$ only, we should look for a value of $\epsilon$ ensuring that $\mathsf{P}_1 \ll \mathsf{P}_2$, so that we may decide on the phase structure quantity threshold using well-known chi-square statistics. 
\begin{figure}[htbp]
\resizebox{\hsize}{!}{
\includegraphics{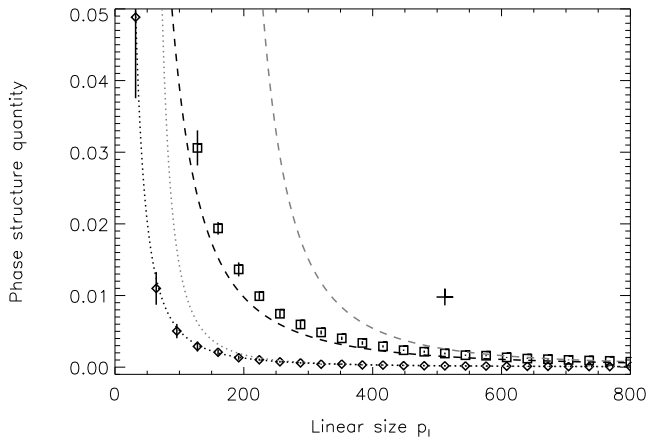}
}
\caption{Evolution of phase structure quantities $\tilde{\mathcal{Q}}(\boldsymbol{e}_x)$ for fractional Brownian motion fields of size $p_l \times p_l$ (implying $p=p_l(p_l-1)$. The symbols represent the mean values of the phase structure quantities for ten realizations of the fractional Brownian fields, while the vertical lines represent the standard deviations. The different symbols correspond to $n=50$ (diamonds) and $n=500$ (squares). The dotted and dashed lines represent the corresponding theoretical upper limits, according to the $\epsilon$-adaptive procedure (grey lines) and the fixed $\epsilon$ procedure (black lines). The cross represents $\tilde{\mathcal{Q}}(\boldsymbol{e}_x)$ in the case of the turbulent column density of Fig.~\ref{fig_4}, for $n=50$.}
\label{fig_7}
\end{figure}

This suggests an ``$\epsilon$-adaptive'' procedure, which is the following: For each $(n,p)$ pair, the value of $\epsilon$ is chosen so that $\mathsf{P}_1$ is small. The quantiles of the chi-square statistics are then used to extract a value of $x$ so that $\mathsf{P}_1 \ll \mathsf{P}_2 \ll 1$. In practice, we took $\mathsf{P}_1=10^{-6}$ and $\mathsf{P}_2=10^{-2}$. In the end, there is a 0.99 probability that the phase structure quantity be less than $x$ with these values of $n$ and $p$, assuming that the underlying distribution is uniform. Conversely, if the measured phase structure quantity $\tilde{\mathcal{Q}}$ is greater than $x$ with these values of $n$ and $p$, then the underlying distribution is most likely non-uniform.

As will be clear later on, this procedure may lead to very conservative upper limits. That is why we may also follow a ``fixed $\epsilon$'' procedure, in which we set $\epsilon$ to a fixed value, say $0.1$, and simply ignore $\mathsf{P}_1$. As in the $\epsilon$-adaptive procedure, we then compute $x$ so that $\mathsf{P}_2=10^{-2}$.


The influence of the number of phase increments $p$ and the number of bins $n$ on the reliability of the phase structure quantity may also be studied numerically. To this end, we have computed a series of two-dimensional fractional Brownian motions of various sizes. The lag vector being fixed, namely $\boldsymbol{\delta}=\boldsymbol{e}_x$, the numerical calculation of the phase structure quantity for these simulations provides us with an estimate of the limit above which one should conclude that phase structure is indeed present in the image. More precisely, we have computed fractional Brownian motion fields of size varying from $32 \times 32$ to $800 \times 800$ pixels\footnote{So that $p$ varies from $32 \times 31$ to $800 \times 799$.}. For each size, ten fields were built and two histograms drawn from the maps of their phase increments, with respectively $n=50$ and $n=500$. For each pair $(n,p)$, we then computed the mean and standard deviation of the ten phase structure quantities associated with the fields. The results are shown on Fig. \ref{fig_7}.

It appears that the computed phase structure quantity increases as the size of the image decreases, and as the number of bins increases. This is interpreted by the fact that the histograms are then less accurate samples of the underlying distributions. The figure also shows theoretical upper limits computed using both procedures described earlier. Unsurprisingly, the $\epsilon$-adaptive upper limits fall above the values found in the numerical simulations, even well above them, which demonstrates the conservativeness of this approach. On the contrary, fixed $\epsilon$ upper limits for $n=50$ match the numerical estimates better, but actually fail when $n=500$. This is due to the fact that, in this latter case, $\mathsf{P}_1 \gtrsim 1$, which makes the upper limit thus computed useless. The position of the phase structure quantity $\tilde{\mathcal{Q}}(\boldsymbol{e}_x)$ for the turbulent column density (Fig.~\ref{fig_4}) in this plot is quantitative evidence that this field does harbour phase structure.

\section{Application to interferometric observations}
\label{sec_atio}

\subsection{Introduction}

In the ideal case, interferometers sample the Fourier transform of observed brightness distributions, and therefore allow direct measurement of the phases of Fourier components. Consequently, it is theoretically possible to have access to phase increments, and to phase structure quantities for the observed fields. Since this can be done in real time, as the Earth's rotation allows for a better sampling of Fourier space \citep{thompson91}, we may look for the minimum observing time required to detect a significant phase structure quantity in the data. This is the topic of this section~\ref{sec_atio}.

Before we carry on, however, we should stress that, in practice, phases measured by the instrument do not directly yield the actual phases of the model brightness distribution. First of all, the antennae are not pointlike receptors, so that brightness distributions are multiplied by a primary beam. In Fourier space, this corresponds to a convolution of the Fourier components by a finite size kernel. Consequently, the phase measured at a given point in the $(u,v)$ plane does not yield the actual phase of the Fourier component of the brightness distribution at that point, but involves all Fourier components within a small neighbourhood. One may circumvent this difficulty by considering mosaicing observation techniques. In short, these amount to imaging large fields by pointing the array towards different directions successively, and ``gluing'' the subfields together \citep{bhatnagar2005}. In image space, this last step is done in such a way that the fall-off due to the primary beam pattern in a subfield is compensated by the rise of the primary beam pattern in the adjacent subfield, so that the effective primary beam is more or less uniform over the large composite field. In Fourier space, these techniques correspond to a finer sampling of the $(u,v)$ plane, effectively reducing the size of the convolution kernel, so that the measured phases are better estimates of the actual phases of Fourier components. We will not discuss this problem any further here, as it should require an extensive study that is not within the scope of this paper, and we assume from now on that antennae can be modelled as pointlike receptors. 

Another point to consider is the fact that measurements in the $(u,v)$ plane are not sampled on a regular grid, and regridding is widely used to allow for Fast Fourier Transforms to be performed \citep{thompson74}. This means that the measured phases end up being associated with a different wavenumber than the one they actually correspond to. In our case, as we use model brightness distributions that are already sampled on a regular grid, the gridding problem can be bypassed\footnote{This is actually only valid in the noise-free case (see~\ref{sec_apn} and~\ref{sec_ext}).}.

Lastly, noise contributions, especially those due to the turbulent fluctuations of the atmosphere, blur the true phase values. This effect will be discussed in detail, in section~\ref{sec_apn}.

\subsection{Simulations of observations}

The instrument simulator used is of the simplest kind, and its parameters, taken to match those of ALMA, are summarized in Table~\ref{tab_2}. The instrument tracks the source as long as it remains above a minimum elevation\footnote{No shadowing of the antennae is taken into account.} of 10$^\circ$, which, given the array's latitude and the source's chosen declination of -20$^\circ$, represents a maximum integration time of 11 hours and 38 minutes. Regarding the number and positions of the antennae on the ground, we have chosen configurations optimized by \citet{boone2001} based on ALMA specifications. For comparison, we have also considered configurations taken from current arrays, such as the Plateau de Bure (PdB) radio-interferometer and the VLA. To make meaningful comparisons, we have used fictitious arrays located at the same geographical coordinates as ALMA, observing the same source. Only the number and positions of the antennae are changed to match the configurations of the PdBI and the VLA. The characteristics of all the arrays used are summarized in Table~\ref{tab_1}. Using the source's apparent movement in the sky and the locations of the antennae on the ground, we obtain ungridded $(u,v)$ covers as functions of integration time.

\begin{table}
\centering
\caption[]{Instrumental parameters}
\vspace{0cm}
\label{tab_2}

\begin{tabular}{|c|c|c|c|}
\hline
Wavelength & Longitude & Latitude & Dump time\\
\hline
1.3 mm & -67.75 $^{\circ}$ &  -23.02 $^{\circ}$ & 10 s\\
\hline
\end{tabular}
\end{table}

\begin{table}
\centering
\caption[]{Array characteristics. For each configuration, the minimum and maximum separation between any two antennae are given.}
\vspace{0cm}
\label{tab_1}

\begin{tabular}{|c||c|c|c|}
\hline
Instrument & ALMA & PdB & VLA\\
\hline
Antenna diameter (m) & 12 & 15 & 25\\
Number of antennae & 60 & 6 & 27\\
A configuration (m) & 19 - 11527 & 32 - 400 & 807 - 37235\\
B configuration (m) & 76 - 3005 & 71 - 331 & 247 - 11314\\
C configuration (m) & 83 - 2303 & 48 - 229 & 79 - 3444\\
D configuration (m) & 43 - 1618 & 24 - 113 & 41 - 1048 \\
E configuration (m) & 34 - 909 & n.a. & n.a. \\
F configuration (m) & 15 - 229 & n.a. & n.a. \\
\hline
\end{tabular}
\end{table}

\begin{figure}[htbp]
\resizebox{0.85\hsize}{!}{
\includegraphics{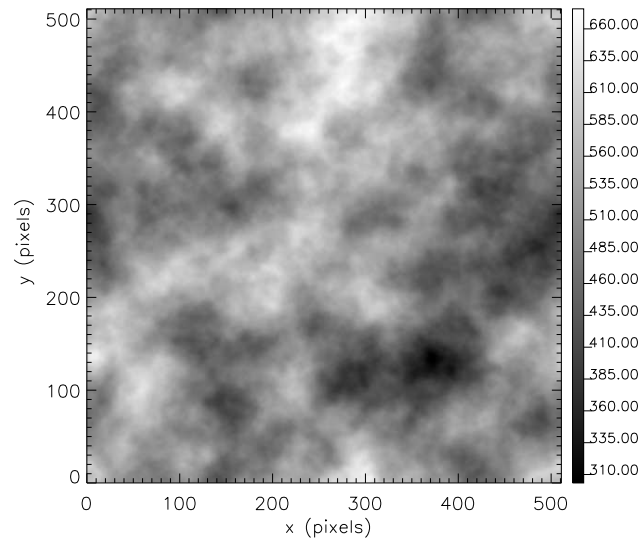}
}
\caption{Synthetic field with power spectrum identical to that of the turbulent column density field (Fig.~\ref{fig_4}), and fully random phases.}
\label{fig_rp}
\end{figure}

\begin{figure}[htbp]
\resizebox{0.95\hsize}{!}{
\includegraphics{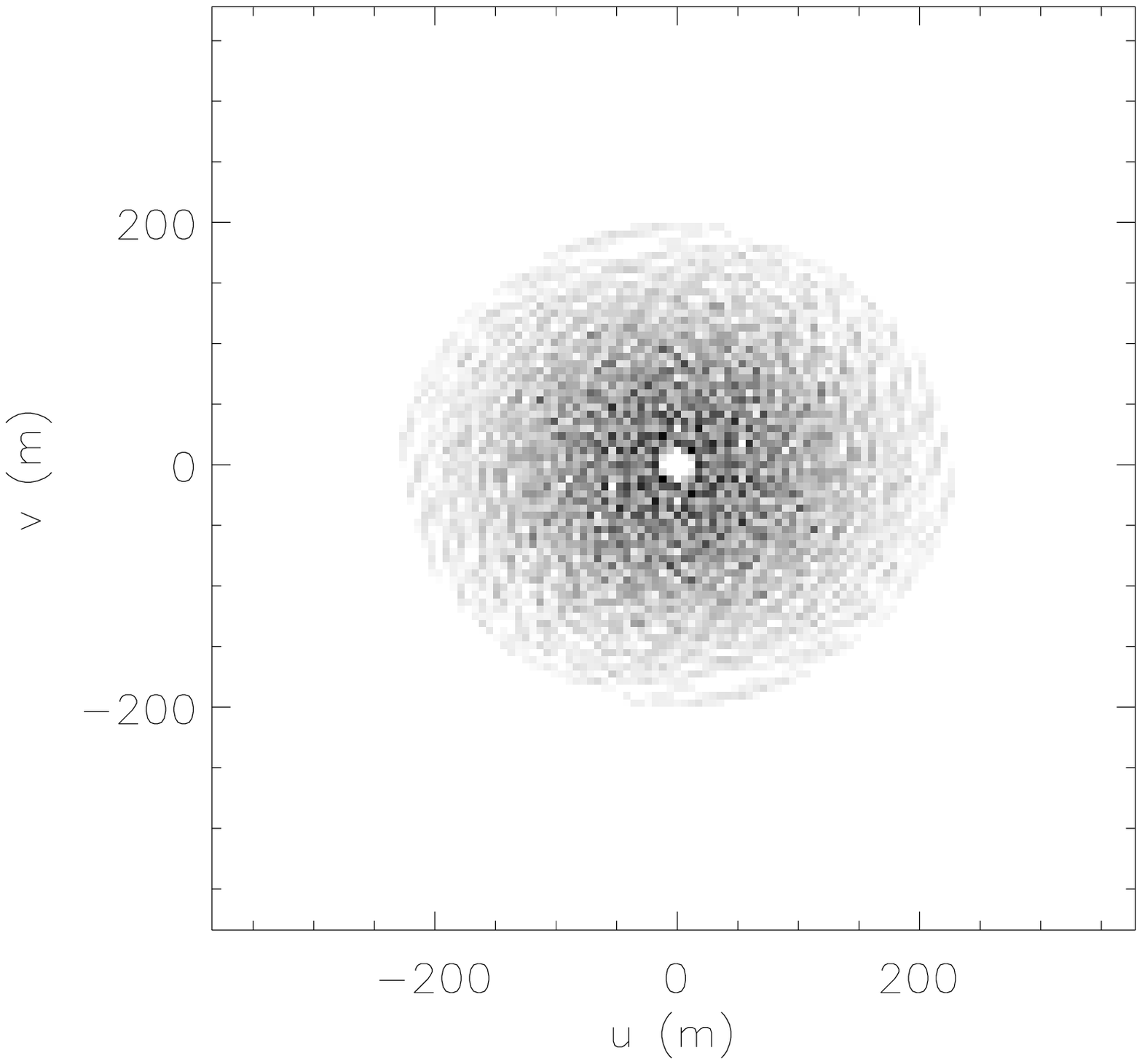}
}
\resizebox{0.95\hsize}{!}{
\includegraphics{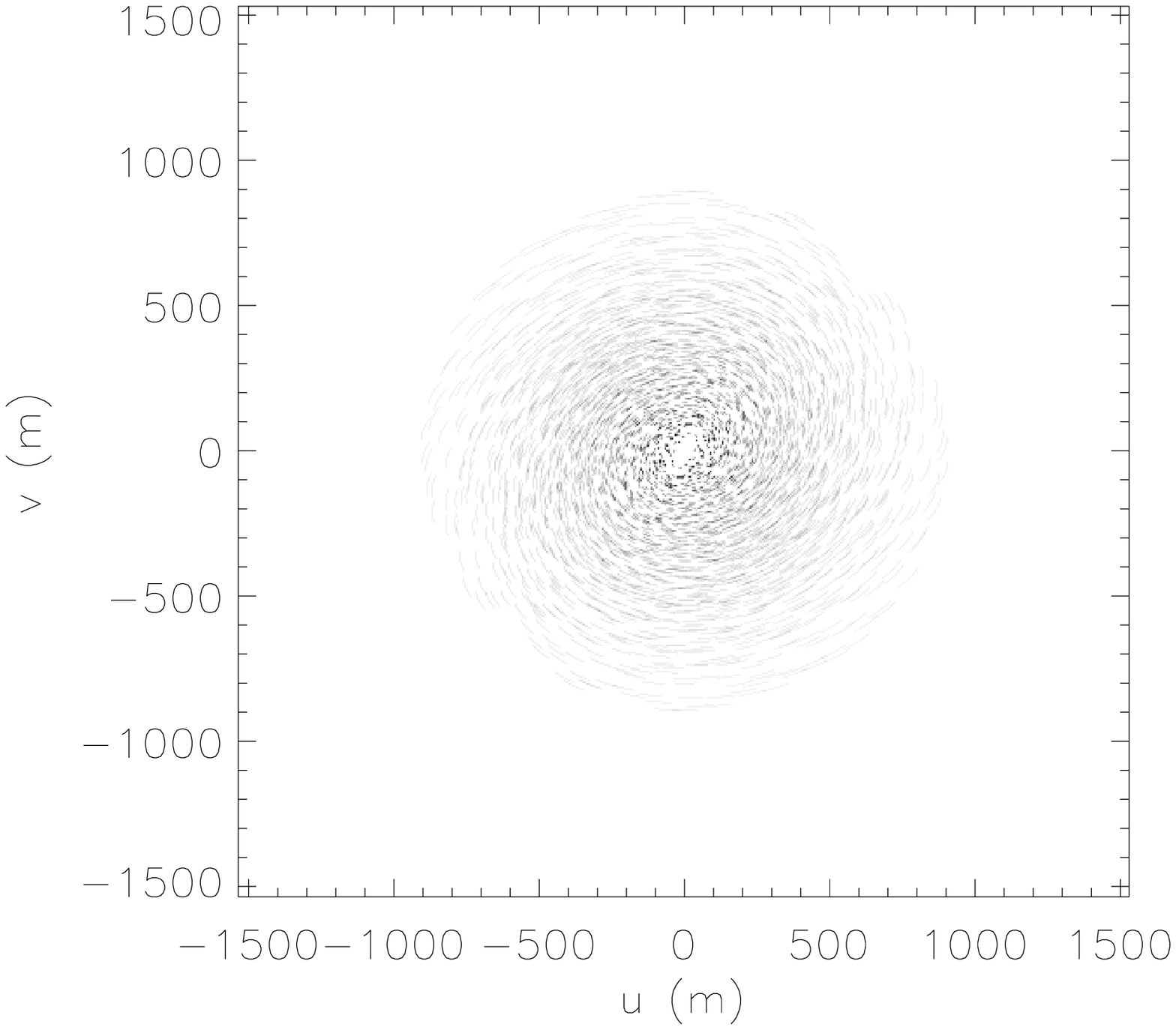}
}
\caption{Gridded $(u,v)$ covers for the F (upper panel) and E (lower panel) configurations of the ALMA instrument, when observing the source during one hour centered on the meridian transit. Pixel sizes are the same in both cases, namely $6 \times 6$ meters, and the F configuration cover has been zoomed. The maximum pixel values are 1726 samples per cell in the F configuration and 470 samples per cell in the E configuration.}
\label{fig_n2}
\end{figure}

As model brightness distribution, we use the turbulent column density shown on Fig.~\ref{fig_4}, which we know harbours phase structure (Fig.~\ref{fig_7}). For comparison purposes, a field with the same power spectrum and random phases is also considered (Fig.~\ref{fig_rp}). Both model fields are $512 \times 512$ images, so $(u,v)$ covers are regridded on a grid of that size, using nearest-neighbour interpolation. The size of the $(u,v)$ cells is chosen to be half the antenna diameter, to satisfy the Nyquist criterion\footnote{This means that the actual pixel size is different for the three arrays considered. We shall not be troubled by this, given that the brightness distributions used are scale-invariant and are not subject to boundary conditions, so that their actual physical extents need not be specified and can therefore be scaled accordingly.}. This limits the size of the maximum baseline that can be considered. Only the F and E configurations of the ALMA instrument, the D configuration of the VLA instrument and all configurations of the PdB instrument fit on $512 \times 512$ grids with the corresponding pixel sizes. These are used in the single-configuration simulations. However, it is possible to consider the more extended configurations, provided the longest baselines are ignored. We did so in the case of multi-configuration observations, as explained in section~\ref{subsec_multi}. As an illustration, Fig.~\ref{fig_n2} shows the gridded covers for the F and E configurations of the ALMA instrument, considering the time period of one hour centered on the source's transit at the meridian.

\subsection{Evolution of the measured phase structure with integration time}

\begin{figure}[htbp]
\resizebox{\hsize}{!}{
\includegraphics{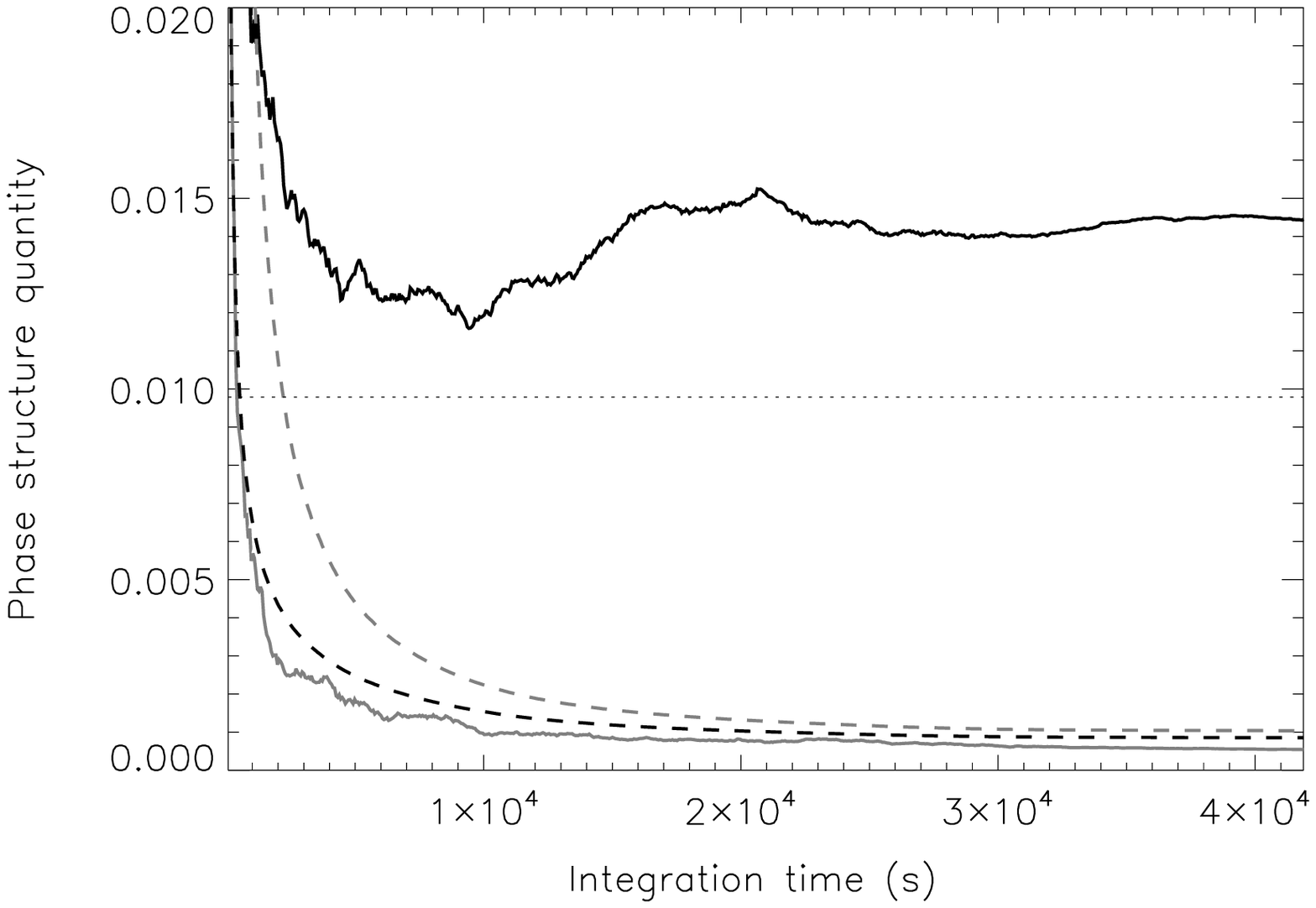}
}
\resizebox{\hsize}{!}{
\includegraphics{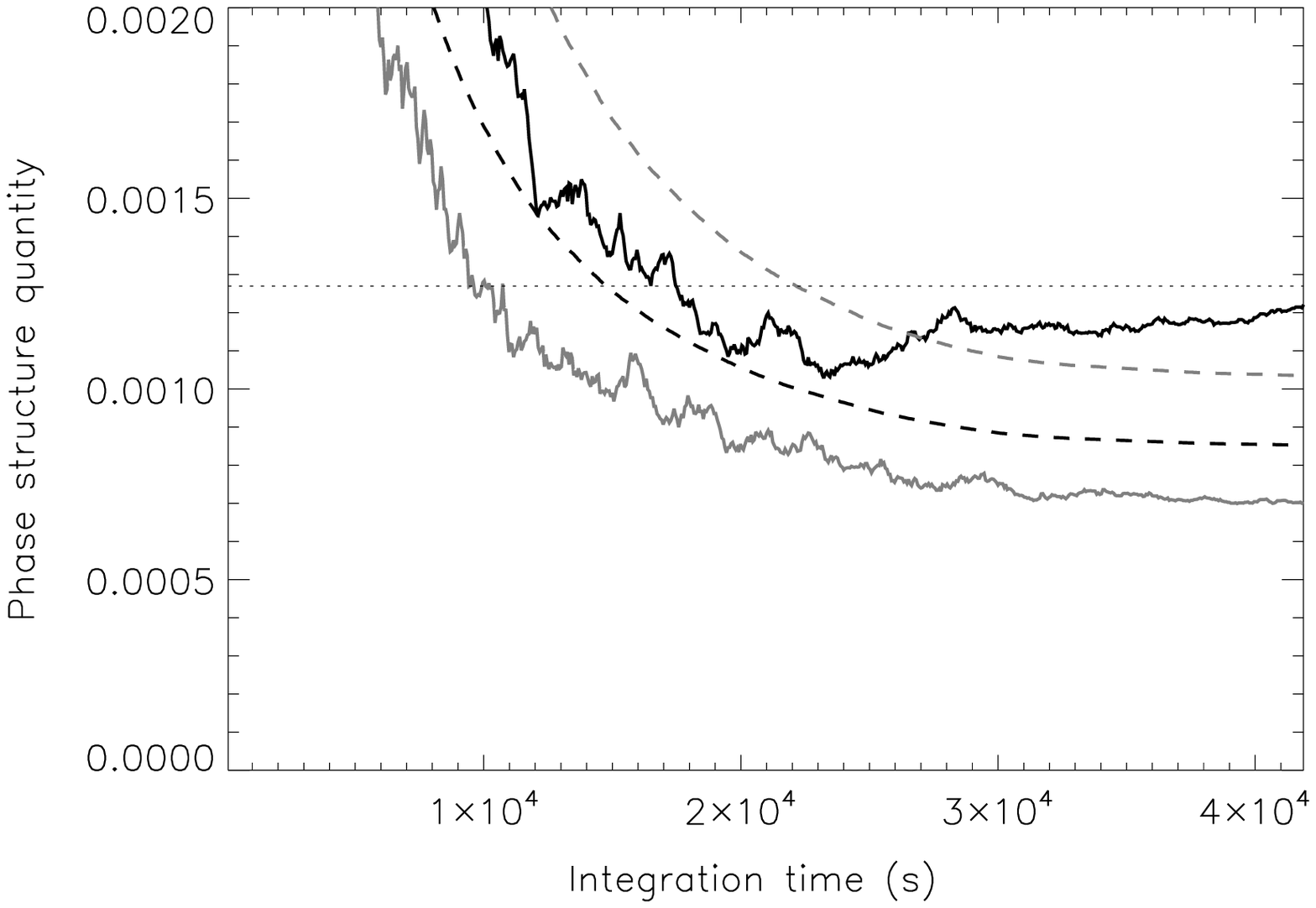}
}
\caption{Evolution of measured phase structure quantities with integration time, for the E configuration of the ALMA instrument. The upper panel corresponds to $\boldsymbol{\delta}=\boldsymbol{e}_x$ and the lower panel to $\boldsymbol{\delta}=\boldsymbol{e}_y$. The black solid lines correspond to the turbulent model brightness distribution (Fig.~\ref{fig_4}), while the grey solid lines correspond to the random-phase brightness distribution (Fig.~\ref{fig_rp}). The dotted lines represent the phase structure quantities of the complete turbulent brightness distribution, for each lag vector, and the dashed lines represent the evolution of theoretical upper limits with integration time, using the fixed procedure (black lines) and the adaptive procedure (grey lines).}
\label{fig_n1}
\end{figure}

\begin{figure}[htbp]
\resizebox{\hsize}{!}{
\includegraphics{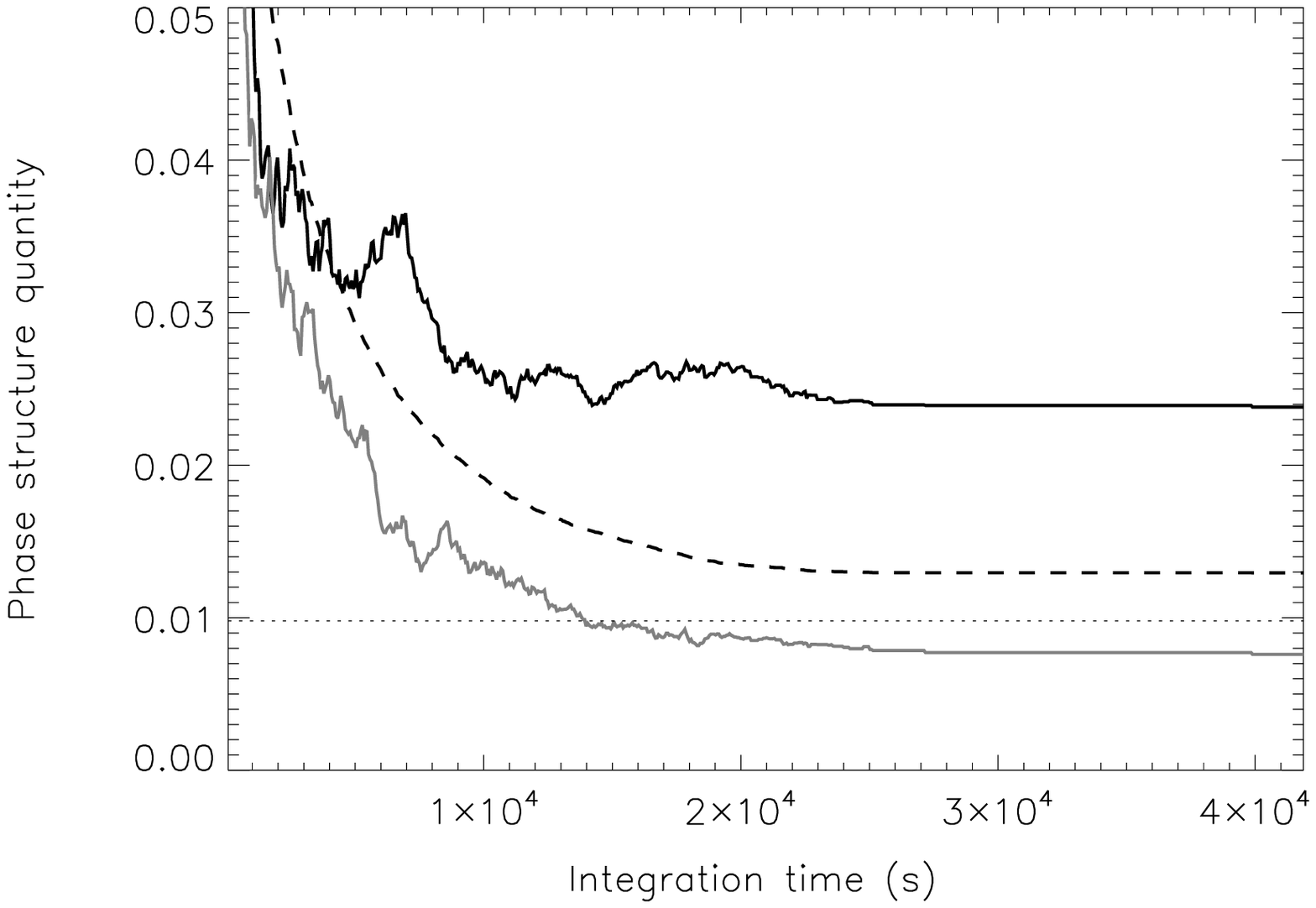}
}
\resizebox{\hsize}{!}{
\includegraphics{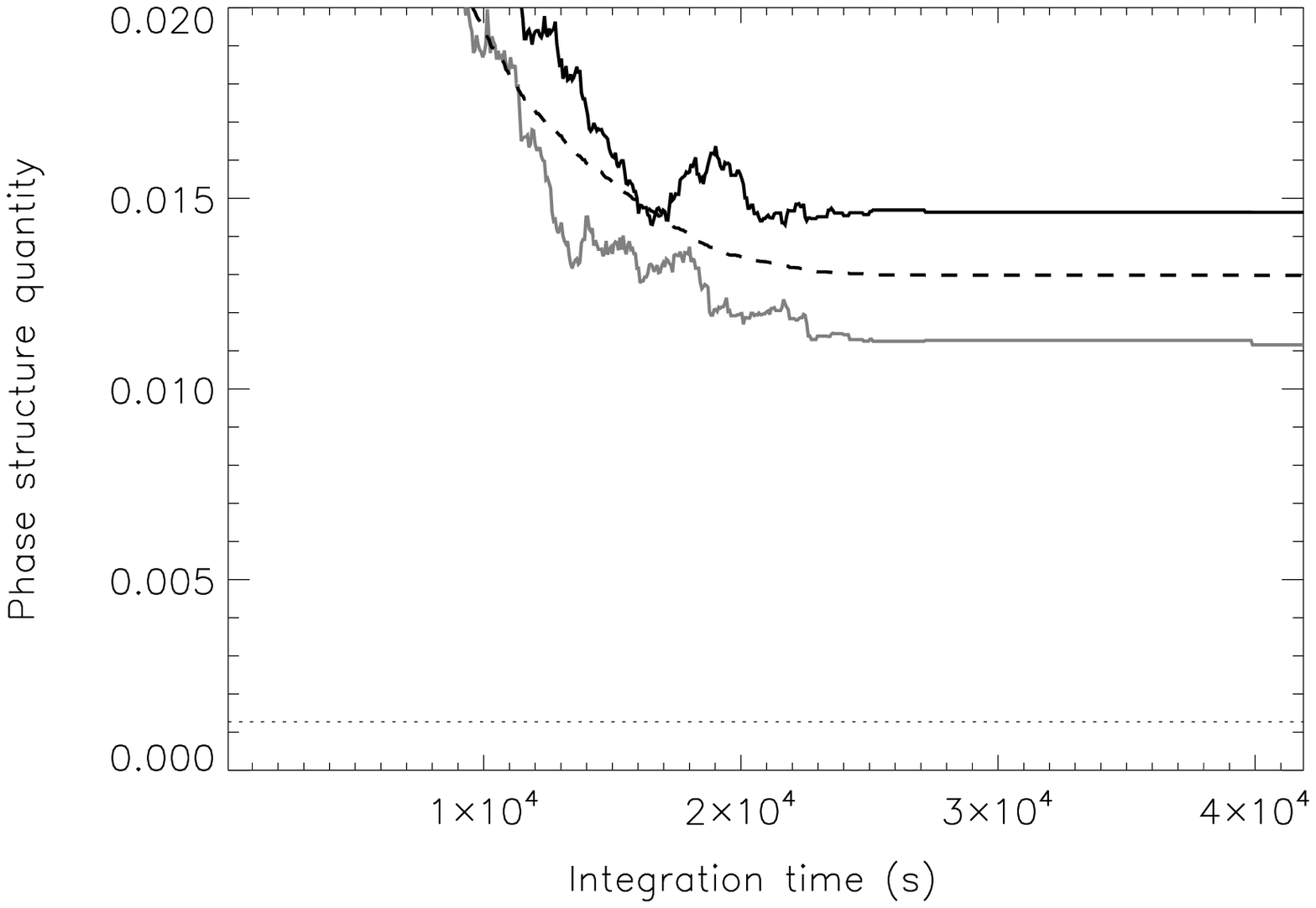}
}
\caption{Same as Fig.~\ref{fig_n1} but for the F configuration of the ALMA instrument. The adaptive procedure's upper limits lie outside of the range of $\tilde{\mathcal{Q}}$ values plotted.}
\label{fig_n3}
\end{figure}

\begin{figure}[htbp]
\resizebox{\hsize}{!}{
\includegraphics{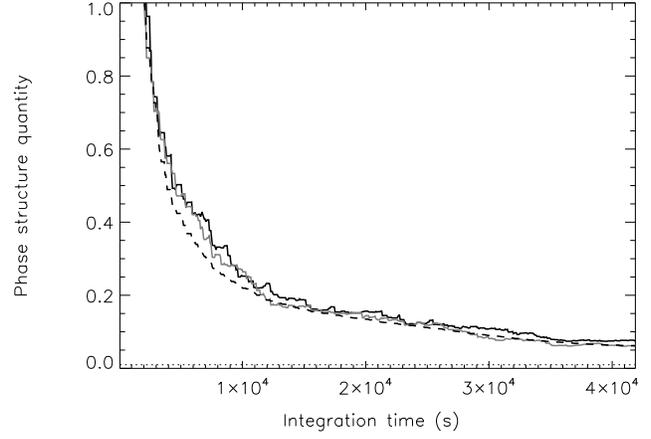}
}
\caption{Same as the top panel of Fig.~\ref{fig_n1}, but for the B configuration of the Plateau de Bure instrument. As with Fig.~\ref{fig_n3}, the adaptive procedure's upper limit lies outside of the range of $\tilde{\mathcal{Q}}$ values plotted.}
\label{fig_n4}
\end{figure}

\begin{figure}[htbp]
\resizebox{\hsize}{!}{
\includegraphics{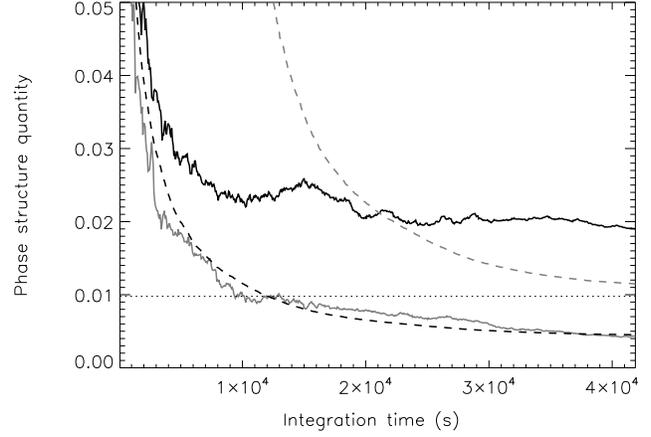}
}
\caption{Same as the top panel of Fig.~\ref{fig_n1}, but for the D configuration of the VLA instrument.}
\label{fig_n5}
\end{figure}

As the observation is carried out, more and more Fourier phases are measured and the number $p$ of phase increments increases, for any given lag vector $\boldsymbol{\delta}$. The question is whether this allows to bring the theoretical and numerical upper limits of Fig.~\ref{fig_7} down sufficiently, below the measured phase structure quantities, to ensure positive detection. To answer this question, Figures~\ref{fig_n1} to~\ref{fig_n5} show the evolution of the measured phase structure quantities as a function of integration time.

Concerning the E configuration of the ALMA instrument (Fig.~\ref{fig_n1}), the conclusions that can be drawn are the following: For $\boldsymbol{\delta}=\boldsymbol{e}_x$, a short integration time of approximately twenty minutes is enough to conclude that phase structure is present in the image, since the measured value then becomes larger than the adaptive procedure's upper limit. On the other hand, the phase structure for $\boldsymbol{\delta}=\boldsymbol{e}_y$ is harder to extract, due to the lower value of $\tilde{\mathcal{Q}}(\boldsymbol{e}_y)$ for the complete field. Approximately 7.5 hours of integration are required to see the measured phase structure quantity rise above the more conservative theoretical upper limit, although the curves for the turbulent field and its random phase version are clearly distinguished on the whole range plotted. When comparing the two panels of Fig.~\ref{fig_n1}, it appears that the measured phase structure quantity for the maximum integration time is larger than the value for the whole field in the case $\boldsymbol{\delta}=\boldsymbol{e}_x$, while it is the opposite for $\boldsymbol{\delta}=\boldsymbol{e}_y$. These discrepancies are due to the limited range of measured spatial frequencies, and are addressed in section~\ref{subsec_multi}. We point out, however, that the approximate constancy of the measured phase structure quantities for $T \gtrsim 3.10^4$ s shows that the $(u,v)$ plane is then close to being fully sampled within this range.

Regarding the F configuration of the ALMA instrument (Fig.~\ref{fig_n3}), its compacity leads to a smaller number of phase increments, which makes phase structure detection all the more difficult. Our best chances lie with the $\boldsymbol{\delta}=\boldsymbol{e}_x$ lag vector (upper panel). In this case, the measured $\tilde{\mathcal{Q}}$ remains constantly below the upper limit set by the adaptive procedure, but becomes larger than that set by the fixed $\epsilon$ procedure, for an integration time of just above one hour. This may be seen as evidence that the phase structure can also be detected with this compact configuration. Whether it should be possible to determine the phase structure quantity for the whole field is uncertain,since it falls below the less conservative upper limit, but above the curve corresponding to a random-phase field. 


Figures~\ref{fig_n4} and~\ref{fig_n5} show the evolution of the measured phase structure quantity $\tilde{\mathcal{Q}}(\boldsymbol{e}_x)$ as a function of integration time, using the B configuration of the Plateau de Bure interferometer and the D configuration of the VLA, respectively. What appears clearly on Fig.~\ref{fig_n4} is that the number of phase increments measured by the Plateau de Bure is insufficient to detect phase structure, as the curves for turbulent and random-phase brightness distributions are indistinguishable from one another. The same conclusion prevails for other configurations of this instrument. On the contrary, the VLA allows such a detection, although it takes a long integration (about 6 hours) to see the phase structure quantity measured emerge from the adaptive procedure's upper limit. Let us note however that the curves for both model brightness distributions start going apart after less than twenty minutes of integration, which should give observers a first hint that phase structure is present in the field. This diagnosis can be performed in real time by drawing random phases for the visibilities as they are measured.

\subsection{Multi-configuration observations}
\label{subsec_multi}

Comparing Figs.~\ref{fig_n1} to~\ref{fig_n5}, the ``best'' situation appears to be the E configuration of the ALMA interferometer, for $\boldsymbol{\delta}=\boldsymbol{e}_x$. Yet, the phase structure quantity obtained after one transit of the source is not the one computed on the whole field. This is due to the fact that only 24\% of the $512 \times 512$ Fourier phases are measured by this configuration. Using more extended configurations, one should be able to recover the Fourier components lying outside the radius covered by the E configuration, and therefore hope to recover the correct value of the phase structure quantity by combining visibilities from multiple configurations. Obviously, this can only be done after the observations have been carried out, unlike what has been considered until now, so that we should rephrase our initial question: what is the minimum integration time necessary in each configuration to ensure that combining the observed visibilities will lead to a ``full'' coverage of the $(u,v)$ plane?\footnote{We should stress that we do not consider a set of integration times $(\tau_1,\ldots,\tau_N)$, one for each of the $N$ configurations, but a single value $\tau$ that applies to each configuration separately.} In this multi-configuration approach, we have to consider the extended configurations that do not fit on a $512 \times 512$ grid with the given pixel size, and simply ignore the visibilities falling outside the grid.

\begin{figure}[htbp]
\resizebox{\hsize}{!}{
\includegraphics{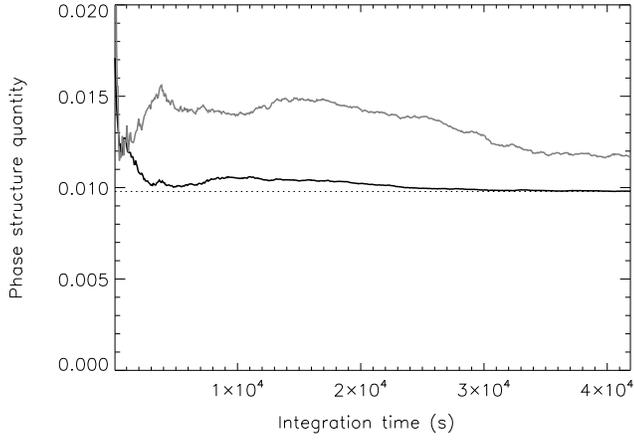}
}
\caption{Evolution of measured phase structure quantity $\tilde{\mathcal{Q}}(\boldsymbol{e}_x)$ with integration time for an observation using all configurations of the instrument in turn. The black solid line corresponds to the six configurations of the ALMA instrument, and the grey solid line to the four configurations of the VLA instrument. The dotted line represents the value of $\tilde{\mathcal{Q}}(\boldsymbol{e}_x)$ for the whole field.}
\label{fig_n7}
\end{figure}

Figure~\ref{fig_n7} shows the evolution of the measured phase structure quantity $\tilde{\mathcal{Q}}(\boldsymbol{e}_x)$ with integration time, using this approach with the ALMA and VLA instruments. In the former case, it appears that the phase structure quantity for the whole field is recovered with an integration time of 9 hours in each configuration, totalling 54 hours of observing time. On the other hand, a full-day integration using the four configurations of the VLA instrument is not sufficient to reach the phase structure quantity of the whole field. This emphasizes the better coverage of the Fourier plane that will be achieved by ALMA.

\subsection{Atmospheric phase noise}
\label{sec_apn}

In the previous subsections, it was assumed that interferometers sample the true Fourier phases of the observed fields. However, it is well known that turbulent motions within the atmosphere above the instrument alter the measured phases. Indeed, these motions cause the amount of water vapor along the line of sight to vary both in time and from one antenna to the other. This results in a delay error and thus a phase error for each baseline. This problem has been addressed thoroughly by~\citet{lay97a,lay97b}, and simulations of the atmospheric fluctuations at the ALMA site of Chajnantor have been performed by \citet{stirling2005}. Here, we have chosen to perform a simulation of the effects of atmospheric phase noise by introducing an atmospheric mask giving the refractivity field $\vartheta(x,y,z)$ above the instrument. In practice, we assumed that this field can be regarded as a 200-m thick layer of frozen Kolmogorov turbulence, that is being transported along the east-west direction at a wind speed of 2 m.s$^{-1}$. Using a spatial resolution of 10 meters, we computed $\vartheta$ as a fractional Brownian motion of size $8500 \times 150 \times 20$ pixels, with spectral index $-11/3$. Integration of $\vartheta$ along the different lines of sight for each antenna as the observation is performed yields phase delays, which are subsequently correlated in order to give the atmospheric phase noise $\phi_a(\alpha,\beta,t)$ for each pair of antennae $(\alpha,\beta)$, at all times $t$. The field $\vartheta$ is normalized so that the rms phase noise $\sigma_0$ for a pair of antennae observing the zenith and separated by a baseline $d=100$ m should be one of a few specific values, namely 15$^\circ$, 45$^\circ$ and 90$^\circ$.

\begin{figure}[htbp]
\resizebox{\hsize}{!}{
\includegraphics{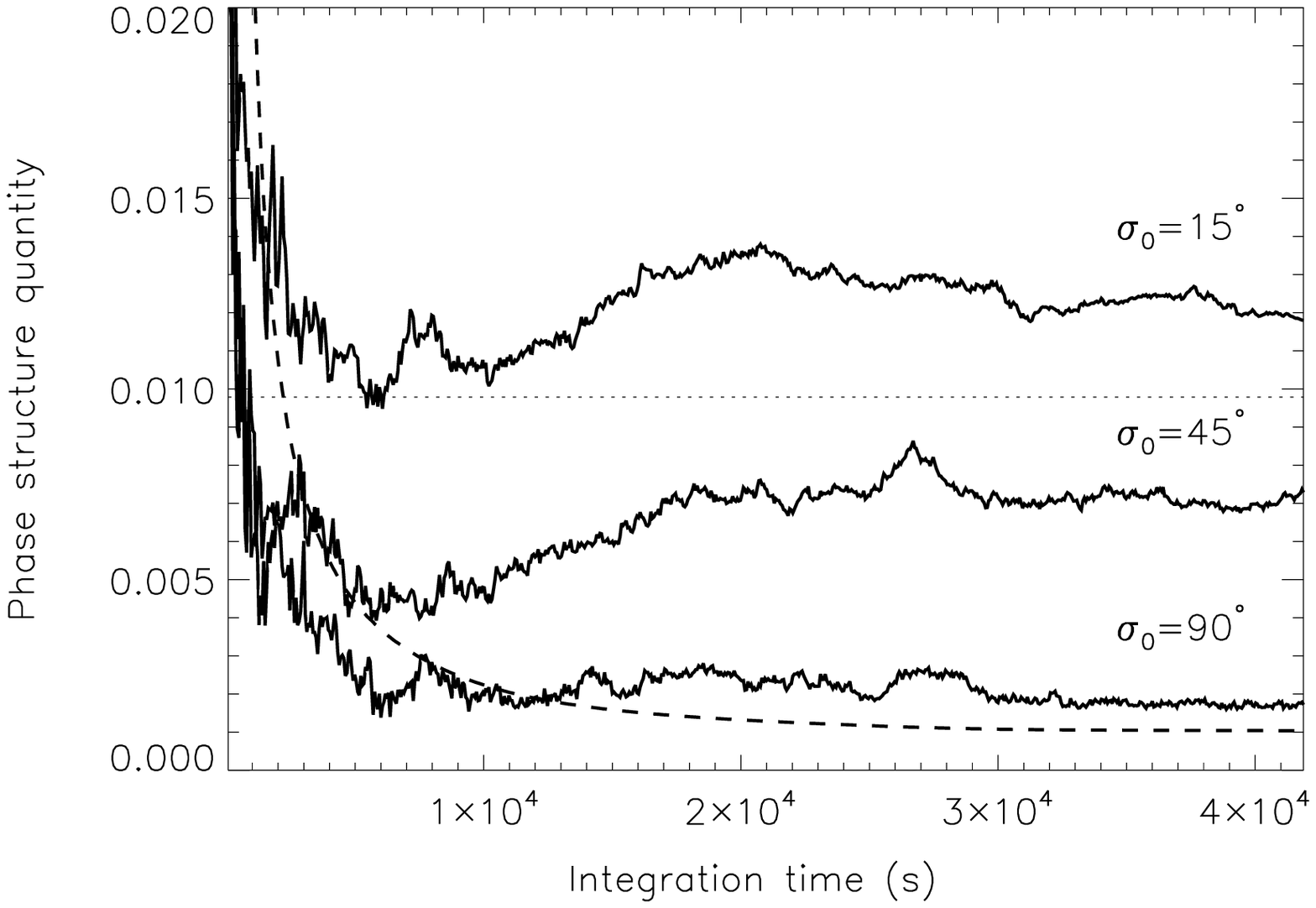}
}
\resizebox{\hsize}{!}{
\includegraphics{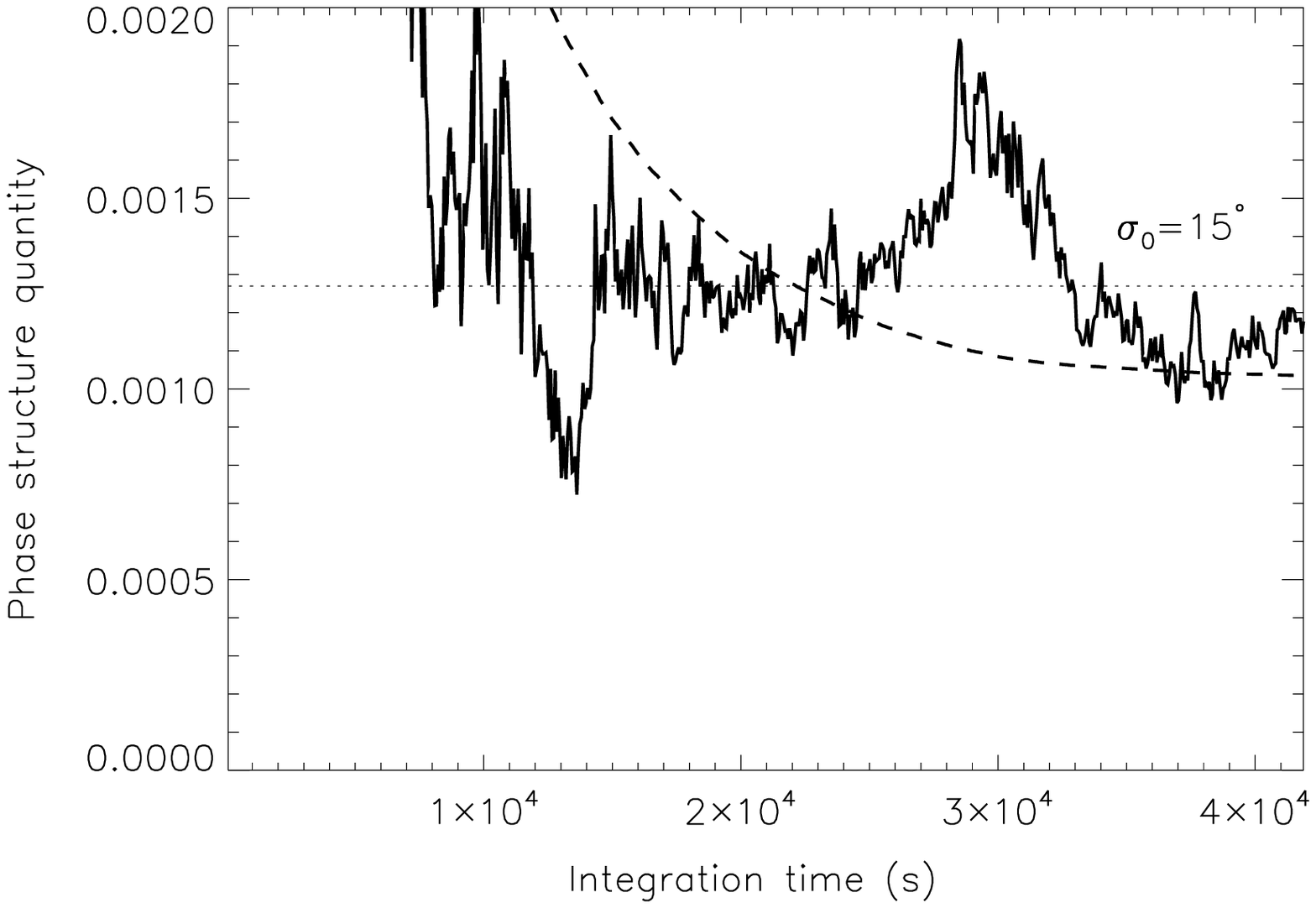}
}
\caption{Evolution of the measured phase structure quantity with integration time, in the presence of atmospheric phase noise (solid lines, with $\sigma_0$ specified next to each curve). The array used is the E configuration of the ALMA instrument, and the top and bottom panels correspond respectively to $\boldsymbol{\delta}=\boldsymbol{e}_x$ and $\boldsymbol{\delta}=\boldsymbol{e}_y$. The dotted lines represent the phase structure quantities for the whole field, and the dashed lines correspond to the $\epsilon$-adaptive upper limits.}
\label{fig_n6}
\end{figure}

At each time step, a number of visibilities fall in each cell of the gridded $(u,v)$ plane, and the observed ``mean visibility'' in the cell $\mathcal{C}_{\boldsymbol{k}}$, centered on wavevector $\boldsymbol{k}$, is given by
\begin{equation}
\label{eq_2}
V'(\boldsymbol{k},t)=\frac{1}{\mathcal{N}(\boldsymbol{k},t)}\sum_{t'\leqslant t}\sum_{(\alpha,\beta,t') \in \mathcal{C}_{\boldsymbol{k}}}V^0(\boldsymbol{k})\exp{[i\phi_a(\alpha,\beta,t')]},
\end{equation}
where $\mathcal{N}(\boldsymbol{k},t)$ is the cumulative number of visibilities within the cell $\mathcal{C}_{\boldsymbol{k}}$ at time $t$, and $V^0$ is the true visibility in this cell. This relation allows for the simulation of the measurement of Fourier phases in the presence of a given amount of atmospheric phase noise. Fig.~\ref{fig_n6} shows the evolution of the measured phase structure quantities $\tilde{\mathcal{Q}}(\boldsymbol{e}_x)$ and $\tilde{\mathcal{Q}}(\boldsymbol{e}_y)$ in the case of the E configuration of the ALMA instrument.

Provided that the value of the phase structure quantity is large enough for the whole field, as is the case for $\boldsymbol{\delta}=\boldsymbol{e}_x$, the presence of phase structure can be easily detected in the presence of a fair amount of atmospheric phase noise. Indeed, even a rms phase fluctuation of 90$^{\circ}$ at 100 m is insufficient to bring the measured phase structure quantity down below the $\epsilon$-adaptive upper limit. In the case of $\boldsymbol{\delta}=\boldsymbol{e}_y$, the conclusion is not so clear-cut: although the $\sigma_0=15^{\circ}$ curve exhibits values larger than the upper limit, it does not remain above this limit after a certain integration time. Faced with such a situation, one should be suspicious of the presence of phase structure in the observed field\footnote{In any case, phase increments corresponding to various lag vectors $\boldsymbol{\delta}$ can be simultaneously computed, in real time, so that for a given observed field, it is enough that there exist one such lag vector for which phase structure detection is feasible.}.

\citet{butler2001} performed measurements of the atmospheric phase fluctuations above the Chajnantor site, using a two-element interferometer observing a 11.198 GHz beacon broadcast by a geostationary satellite positioned 35$^{\circ}$ above the horizon. Using the scaling relation~\citep{lay97a,stirling2005}
\begin{equation}
\label{eq_1}
\sigma_\phi(d,\lambda,\zeta) =\sigma_0 \left(\frac{d}{100~\mathrm{m}}\right)^{5/6}\left(\frac{1.3~\mathrm{mm}}{\lambda}\right)\left(\cos{\zeta}\right)^{-3/4},
\end{equation}
giving the rms phase delay as a function of the baseline length $d$, the wavelength $\lambda$ and the elevation angle $\zeta$, their results translate to a noise level showing diurnal as well as seasonal variations going from $\sigma_0 \sim 14^{\circ}$ to $\sigma_0 \sim 57^{\circ}$. Consequently, the phase structure quantity for $\boldsymbol{\delta}=\boldsymbol{e}_x$ will undoubtedly be detected without any phase correction, although the use of dedicated water vapor radiometers, as is planned for ALMA, should allow for an effective decrease of the atmospheric phase noise by a substantial factor \citep{lay97b}, making possible the measurement of the actual phase structure quantity for the whole field, using multiple configurations.

\subsection{Extensions}
\label{sec_ext}

It should be stressed that the approach used here is not to be considered optimal, but only as evidence that ALMA will be able to detect phase structure in the presence of atmospheric phase noise. Indeed, consider two tracks in the $(u,v)$ plane, corresponding to two different antenna pairs, going through the same grid cell. In the above approach, the measured phases in this cell may be very different for one antenna pair and the other, since these sample different lines of sight through the atmospheric mask. As a result, the estimation of phase through the ``averaged visibility'' of Eq.~(\ref{eq_2}) is badly contaminated. A more elaborate use of phase information would therefore have to be baseline-based. Keeping track of the phase measured by each baseline as a function of time, and computing phase increments along the baseline's track should markedly reduce contamination by atmospheric phase noise. An important point in this approach, which is currently under study, is that the lag vector $\boldsymbol{\delta}$ is no longer a control parameter, but a function of time and of the baseline. 

Conversely, the consideration of the radial evolution of phase structure quantities leads to another possible extension of this work, which is the inclusion of the kinematic dimension. Velocity information is indeed accessible with high spectral resolution receivers such as those that will be used for ALMA (4096 spectral channels over 16 GHz bandwidths). Consequently, Fourier phase analysis applied to individual channel maps may prove a valuable tool for assessing the three-dimensional structure of velocity fields. We may for instance wish to compare phase structure quantities found across line profiles, and see if values found for individual channels are greater than those found for integrated emission maps, which is likely to be the case, as they present higher contrasts.

\section{Summary and conclusions}
\label{sec_conc}

In this paper, we have addressed the ability of radio-interferometers to detect and recover the information contained in the Fourier-spatial distribution of phases, which was previously shown to store a vast amount of information about the structure of images. The PDF of phase increments and the related concept of phase entropy were introduced in this perspective. We ourselves have used the phase structure quantity $\mathcal{Q}$, which is a minor modification of phase entropy leading to $\mathcal{Q}=0$ for fields with purely random phases.

Our main conclusion is that the dynamical range of spatial frequencies observed by the instrument is the key parameter allowing detection and measurement of phase structure.

Using a turbulent model brightness distribution and instrumental configurations based on the characteristics of the future ALMA interferometer and of two existing arrays (VLA and Plateau de Bure), we have assessed the minimum integration time required by each configuration to have a significant detection of phase structure in the observed field. In the most conservative assessment, it appears that for a whole-field phase structure quantity $\mathcal{Q} \simeq 10^{-2}$, detection is achieved with a twenty minute integration in the ALMA E configuration (baselines going from 34 m to 909 m), or with a six hour integration in the VLA D configuration, but is not achieved by any other instrumental configuration tested\footnote{It should be reminded that the more extended configurations have not been used in this single-configuration approach.}. With a whole-field phase structure quantity $\mathcal{Q} \simeq 10^{-3}$, certain detection can only be achieved using the ALMA E configuration, in which case it takes about 7.5 hours of integration. 

However, less conservative criteria allow for early hints at the presence of phase structure in the observed field. Indeed, by drawing random phases for the visibilities in real time, it is possible to compare the evolution of the phase structure quantity for the observed field to that for a random-phase field, and check if they start going apart at some point. This is the case for all ALMA and VLA configurations, with whole-field phase structure quantities $\mathcal{Q} \simeq 10^{-2}$ and $\mathcal{Q} \simeq 10^{-3}$, but not for any of the Plateau de Bure configurations.

Regarding the possibility to recover the actual values of the phase structure quantity for the complete field, only multi-configuration observations with the ALMA instrument seem to allow for it, and it takes 9 hours in each of the 6 configurations to achieve this. 

Finally, we have studied the influence of atmospheric phase noise on the single-configuration observations, using the E configuration of ALMA and whole-field $\mathcal{Q}\simeq 10^{-2}$. The maximum rms phase fluctuations that can be allowed without completely washing out the actual phase structure lie well above the typical range of variations for the Chajnantor site. Consequently, the use of water vapor radiometers to correct for the atmospheric phase fluctuations does not appear as a necessary feature of the ALMA array in this respect, although it should allow for a more accurate determination of the actual phase structure quantity in the multiconfiguration scheme.

Possible extensions to this work include the study of phase increments along the baseline tracks, which should considerably reduce the effects of atmospheric phase noise, and the evolution of phase structure with frequency in high spectral-resolution observations of line sources.

\bibliographystyle{aa}
\bibliography{all}

\appendix

\section{Theoretical upper limits}


Independently of the underlying distribution, the parameters which have an influence on the histograms of phase increments are the number of samples $p$ and the number of bins $n$. Assume then that $p$ phase increments are drawn from a uniform distribution on $\mathcal{A}=[-\pi,\pi]$ and distributed over $n$ intervals $\mathcal{A}_i$ of equal length, thus yielding a $n$-binned histogram $h$. The phase structure quantity $\tilde{\mathcal{Q}}$ associated with $h$ is nonzero due to the sampling noise. We therefore wish to obtain an upper limit to the probability $\mathsf{P}\left(\{\tilde{\mathcal{Q}} > x\}\right)$, as a function of $x$. The histogram values $\left\{h_i\right\}$ are normalized according to 
\begin{equation*}
\frac{2\pi}{n}\sum_{i=1}^n h_i=1.
\end{equation*}
Let us then define the functions
\begin{equation*}
s_0(x)=\frac{1}{2\pi}\boldsymbol{1}_{\mathcal{A}} \quad\! \textrm{and} \quad\! s(x)=\sum_{i=1}^nh_i\boldsymbol{1}_{\mathcal{A}_i},
\end{equation*}
where the symbol $\boldsymbol{1}_{[a,b]}$ stands for the function which is equal to one on the interval $[a,b]$ and zero outside. The phase structure quantity associated with $h$ may then be written as
\begin{equation*}
\tilde{\mathcal{Q}}=-\int\nolimits_{-\pi}^{\pi}s_0(x)\ln{[s_0(x)]}\ud x+ \int\nolimits_{-\pi}^{\pi}s(x)\ln{[s(x)]}\ud x.
\end{equation*}
Now, since $s_0$ is actually a constant on $[-\pi,\pi]$, and since both $s$ and $s_0$ are normalized to unity, it is straightforward to obtain
\begin{equation*}
\tilde{\mathcal{Q}}=\int\nolimits_{-\pi}^{\pi}s(x)\ln{\left[\frac{s(x)}{s_0(x)}\right]}\ud x,
\end{equation*}
which shows that the phase structure quantity may be interpreted as a K\"ullback pseudo-distance of $s$ to $s_0$. The method of \citet{castellan2000} suggests to find an upper limit to $\mathsf{P}\left(\{\tilde{\mathcal{Q}} > x\}\right)$ by treating separately the cases of {\it regular} and {\it extraordinary} histograms, the latter being when the histogram presents an unusually large or unusually low value. Let us then define, for any $\epsilon>0$, the event
\begin{equation*}
\Omega_{\epsilon}=\left\{\exists i; |x_i-r|>\epsilon r\right\} \quad\! \textrm{with} \quad\! x_i=\frac{2\pi}{n}h_i \quad\!\textrm{and}\quad\!  r=\frac{1}{n}.
\end{equation*}
The event $\Omega_{\epsilon}$ is precisely the occurrence of an extraordinary histogram, and it depends on the real number $\epsilon$. For instance, if $\epsilon=0.1$, this event occurs if one of the histogram values deviates from the uniform value $1/(2\pi)$ by more than ten percent. The value of $x_i$ may here be interpreted as the mean number of successful events in a series of $p$ Bernoulli trials, the event in question, whose probability is $r$, being that a phase increment belongs to interval $\mathcal{A}_i$. The usual values of $p$ and $n$ (in our case $p=512^2$ and $n=50$) make it reasonable to assume that the central limit theorem applies. We may then write
\begin{equation*}
\mathsf{P}(\Omega_{\epsilon}) \leqslant n\mathsf{P}\left(\left\{|x_i-r|>\epsilon r\right\}\right) \approx n\left[1-\mathrm{Erf}(x_{\epsilon})\right],
\end{equation*}
introducing the value of the error function 
\begin{equation*}
\mathrm{Erf}(x)=\frac{2}{\sqrt{\pi}}\int\nolimits_{0}^{x}e^{-t^2}\ud t \quad \textrm{at} \quad x_{\epsilon}=\sqrt{\frac{p}{2(n-1)}}\epsilon.
\end{equation*}
Numerically, for $p=512^2$ and $n=50$, the probability that there exist an interval $\mathcal{A}_i$ containing a number of increments different from the theoretical value by over five percent $(\epsilon=5\,10^{-2})$ is less than about 0.012, and it falls below $10^{-11}$ for a ten percent discrepancy. In this case, it will be possible to neglect the contribution of extraordinary histograms.

Regarding the regular histograms, for which $\Omega_\epsilon$ does not occur, we have the following result, due to \citet{castellan2000},
\begin{equation*}
\int\nolimits_{-\pi}^{\pi}\!\!\mathrm{inf}(s,s_0)\left[\ln{\left(\frac{s}{s_0}\right)}\right]^2\ud x \leqslant 2\tilde{\mathcal{Q}} \leqslant \int\nolimits_{-\pi}^{\pi}\!\!\mathrm{sup}(s,s_0)\left[\ln{\left(\frac{s}{s_0}\right)}\right]^2\ud x.
\end{equation*}
For the regular histograms considered here, we obviously have $(1-\epsilon)s_0 \leqslant \mathrm{inf}(s,s_0) \leqslant \mathrm{sup}(s,s_0)\leqslant (1+\epsilon)s_0$, so that, using
\begin{equation*}
\frac{1}{(1+\epsilon)^2}\frac{\chi^2}{p} \leqslant \int\nolimits_{-\pi}^{\pi}\!\! s_0\left[\ln{\left(\frac{s}{s_0}\right)}\right]^2\ud x \leqslant \frac{1}{(1-\epsilon)^2}\frac{\chi^2}{p},
\end{equation*}
which were obtained by \citet{castellan2000}, we conclude that the phase structure quantity is equivalent to the $\chi^2$ statistics of degree\footnote{The degree is $n-1$ and not $n$ because of the constraint that $s$ should be normalized to unity.} $n-1$, for regular histograms,
\begin{equation*}
a_{\epsilon}\frac{\chi^2}{p} \leqslant \tilde{\mathcal{Q}} \leqslant b_{\epsilon}\frac{\chi^2}{p} \quad\!\! \textrm{with} \quad\!\! a_{\epsilon}=\frac{1-\epsilon}{2(1+\epsilon)^2} \,\, \textrm{and} \,\, b_{\epsilon}=\frac{1+\epsilon}{2(1-\epsilon)^2}.
\end{equation*}
For regular histograms, the probability we are concerned with is therefore subject to the inequality
\begin{equation*}
\mathsf{P}\left(\{\tilde{\mathcal{Q}} > x\}\right) \leqslant \mathsf{P}\left(\left\{b_{\epsilon}\frac{\chi^2}{p} > x\right\}\right)=\mathsf{P}\left(\left\{\chi^2 >\frac{2(1-\epsilon)^2px}{1+\epsilon}\right\}\right),
\end{equation*}
and taking into account both regular and extraordinary histograms, we come up with the upper limit given in the main body of the paper.

\end{document}